\documentclass[aps,prapp,preprint,floatfix,showpacs,citeautoscript,longbibliography]{revtex4-1}
\usepackage{relsize}
\usepackage{graphicx}
\usepackage{float}
\usepackage{bm}
\usepackage{epsfig,booktabs}
\usepackage{times}
\usepackage{wrapfig}
\usepackage{color}
\usepackage{amsmath,amssymb,amsfonts,amsthm}
\usepackage{nccmath}
\usepackage[T1]{fontenc}
\pdfoutput=1

\begin{document}
	
\title{Effects of Disorder on Thermoelectric Properties of Semiconducting Polymers}

\author{Meenakshi Upadhyaya}
\affiliation{Department of Electrical and Computer Engineering, University of Massachusetts Amherst, Amherst, MA 01003-9292, USA}
\author{Connor J. Boyle}
\author{Dhandapani Venkataraman}
\affiliation{Department of Chemistry, University of Massachusetts Amherst, Amherst, MA 01003-9292, USA}
\author{Zlatan Aksamija} \email{zlatana@engin.umass.edu}
\affiliation{Department of Electrical and Computer Engineering, University of Massachusetts Amherst, Amherst, MA 01003-9292, USA}


\begin{abstract}
Organic materials have attracted recent interest as thermoelectric (TE) converters due to their low cost and ease of fabrication. We examine the effects of disorder on the TE properties of semiconducting polymers based on the Gaussian disorder model (GDM) for site energies while employing Pauli's master equation approach to model hopping between localized sites. Our model is in good agreement with experimental results and a useful tool to study hopping transport. We show that stronger overlap between sites can improve the electrical conductivity without adversely affecting the Seebeck coefficient. We find that positional disorder aids the formation of new conduction paths with an increased probability of carriers in high energy sites, leading to an increase in electrical conductivity while leaving the Seebeck unchanged. On the other hand, energetic disorder leads to increased energy gaps between sites, hindering transport. This adversely affects conductivity while only slightly increasing Seebeck and results in lower TE power factors. Furthermore, positional correlation primarily affects conductivity, while correlation in site energies has no effect on TE properties of polymers. Our results also show that the Lorenz number increases with Seebeck coefficient, largely deviating from the Sommerfeld value, in agreement with experiments and in contrast to band conductors. We conclude that reducing energetic disorder and positional correlation, while increasing positional disorder can lead to higher TE power factors.	
\end{abstract}

\flushbottom
\maketitle

\thispagestyle{empty}
	
\section{Introduction}
	
The increased interest in electricity generation from waste heat and natural sources \cite{BellSci08}, as well as large-scale Peltier coolers \cite{SharpIEEE06}, driven by the need for alternative energy sources, has led to extensive research on more cost-effective and energy-efficient TE materials. TE conversion efficiency of a given material is governed by its dimensionless figure of merit $	ZT=S^2\sigma T \left/\kappa \right.$, 
	%\begin{equation} \label{eq:zt}
	%	ZT=\frac{S^2\sigma T} {\kappa}=\frac{S^2}{L}\frac{\kappa_e}{\kappa_e+\kappa_l}
	%\end{equation}
where S is the Seebeck coefficient, $\sigma$ is the electrical conductivity, T is the absolute temperature, and $\kappa=\kappa_{l}+\kappa_{e}$ is the total thermal conductivity composed of electronic ($\kappa_e$) and lattice contributions ($\kappa_l$). The electronic component of thermal conductivity is related to the electrical conductivity through the Wiedemann-Franz law $\kappa_e=L \sigma T$ \cite{Franz53}, where L is the Lorenz number. Most doped semiconductors have a high $\sigma$ and a moderate Seebeck coefficient, but ZT is still limited by their high lattice thermal conductivity. There has been much research effort to reduce the thermal conductivity by nanostructuring \cite{GonzalezRev13}, embedding nanoparticles in the semiconductor \cite{KimPRL06} and surface roughness \cite{HochbaumNAT08,MartinPRL09,MaurerAPL15}, which increase scattering and limit phonon transport. 
	
Organic materials offer several advantages as they have an inherently low thermal conductivity on account of their disordered structure and thus do not require further processing such as nanostructuring to reduce it. Thermal conductivity in $\pi$-conjugated polymers such as poly(3-hexylthiophene) (P3HT), polypyrrole (PPY), polycarbazoles, polythiophenes, polyaniline (PANI) and poly(3,4-ethylenedioxythiophene) (PEDOT)\cite{GonzalezRev13, RothJCP91, ZhangSM10, DubeyJPP11} is at least an order of magnitude lower than inorganic compounds, with lattice thermal conductivity typically $<$1 Wm$^{-1}$K$^{-1}$ \cite{KimNMat13,SinghIEEE01,LiuACS15}.  They can be chemically oxidized or reduced (doped) to induce high levels of carrier density and thus improve their electrical conductivity \cite{HeegerPRL78,KimNMat13}. They have many additional advantages, such as low cost due to inexpensive fabrication methods and large-area production, solubility in common solvents, and solution processing \cite{RogersAdvMat99}. Thin films can be easily manufactured using techniques such as solution casting \cite{RogersAdvMat99}, vacuum evaporation \cite{HorowitzSSC89}. and  printing technologies \cite{PedeMSE98}. They are also potentially disposable and have lower potential for negative environmental impact \cite{SinghIEEE01}, which makes them a very attractive choice for commercially viable TE applications \cite{CulebrasRev14,MingEnerEnvSc13,DubeyJPP11}.
%missing reference HuPT07

Doping polymers in order to improve the electrical conductivity has the undesirable effect of significantly reducing their Seebeck coefficient to a range in the order of tens of $\mu$VK$^{-1}$.\cite{Sun10} Therefore, a long-standing problem in TEs has been to effectively decouple conductivity from the Seebeck coefficient and control them independently. Nonetheless, doping polymer blends with a minor additive component can result in a simultaneous increase of conductivity and Seebeck coefficient \cite{Sun10}. Also, polymer-CNT composites and organic-inorganic composite materials have been shown to possess higher ZT values due to increased electrical conductivity of these materials \cite{YuACS11, GonzalezRev13, TongaAMI17}. Further improvements by mixing with graphite/graphene, carbon nanotubes, or inorganic TE nanoparticles have also been observed.\cite{CulebrasRev14} Doping PEDOT with counter ions has shown to increase the TE efficiency up to two orders of magnitude depending on the counter-ion size \cite{CulebrasJChem14}. Recently, PEDOT:Tos with enhanced Seebeck coefficient of 210 $\mu$V, electrical conductivity of 70 Scm$^{-1}$ with a resulting ZT$\sim$0.25 has been reported \cite{BubnovaNmat11}, making them a viable alternative to inorganic materials. % missing citation GaoCMS06
	  
Polymer systems do not possess the continuous order found in their inorganic counterparts; they are inherently disordered and charge transport can be described as a hopping process \cite{PautmeierSynMat90, BasslerJCP91,TesslerAdvMat09}. Simulation is a powerful tool to understand transport in these materials and investigate the effect of disorder on the TE performance of these materials. One of the most prominent phenomenological models is based on variable-range hopping (VRH) of electrons between the polymer chains and called the Gaussian Disorder Model (GDM) \cite{BasslerJCP91}. This model has been widely used to study charge transport in polymers and polymer based devices \cite{BasslerPSS93,PasveerPRL05,GrovesJCP08,FishchukPRB07,VandelholstCh2}. More recently, charge transport in organic systems have also been analyzed by combining ab initio calculations with classical molecular dynamics (MD) and kinetic Monte Carlo techniques \cite{KirkpatrickPRL07,NelsonJCP10,WangNL09}. In the work by Mendels and Tessler \cite{TesslerPCL14}, positional disorder was implemented as spatial variations without including the variation in orbital overlap. However, the TE performance of these materials and the combined effects of disorder and correlation on TE transport, especially within a hopping model, has not been fully explored. Also, there have been few studies to determine the Lorenz number and its relationship with Seebeck coefficient in semiconducting polymers. 
	
We investigate the TE properties of disordered organic semiconductors employing a model based on the GDM and use Pauli's master equation (PME) approach to calculate site occupational probabilities. Most studies based on the GDM and its variants compute the hopping rate between adjacent sites using the Miller-Abrahams rate equation. We present a comparison of the Miller-Abrahams hopping rate with the Marcus hopping rate, which considers the additional energy penalty to hopping due to polaronic binding. We explore the effect of various manifestations of disorder, including positional disorder, energetic disorder, as well as correlation in both energy and  wave-function overlap distributions, on the electrical conductivity, Seebeck coefficient and Lorenz number. We find that the overlap has an enormous impact on electrical conductivity whereas spatial variations have negligible effect. We find that positional disorder aids the formation of new conduction paths with an increased probability of carriers in high energy sites leading to increase in electrical conductivity. In contrast, energetic disorder leads to increased energy gap between sites hindering transport and adversely affects conductivity, however, the increased energy gap also leads to a lower average site energy and a small increase in Seebeck coefficient. Consequently, positional correlation negatively affects conductivity, while correlation in energy has no effect on TE properties of polymers. 
	 
Experimental studies on iodine-doped polyacetylene \cite{MermilliodJPhys80} and PEDOT:PSS \cite{LiuACS15} have shown that the Wiedemann-Franz law holds and the Lorenz number is close to the Sommerfeld value. However, Weathers et al. \cite{WeathersAdvMat15} showed the electronic contribution to thermal conductivity was higher than previously reported, consistent with a large Lorenz number, while Lu et al. \cite{NianduanJAP16} reported a large deviation from the Wiedemann-Franz law under the effect of temperature, carrier concentration, energetic disorder, and electric field. We find that Lorenz number increases with Seebeck coefficient, largely deviating from the Sommerfeld value, and it increases further with increasing positional and energetic disorder. The presence of disorder leads to inherently different transport and the design of efficient polymer TEs requires consideration of both positional and energetic disorder, as well as the Lorenz number, in addition to optimizing the doping concentration.
	   
\section{Polymer Theory}

\begin{figure}
	\centering
	\includegraphics[width=0.8\textwidth]{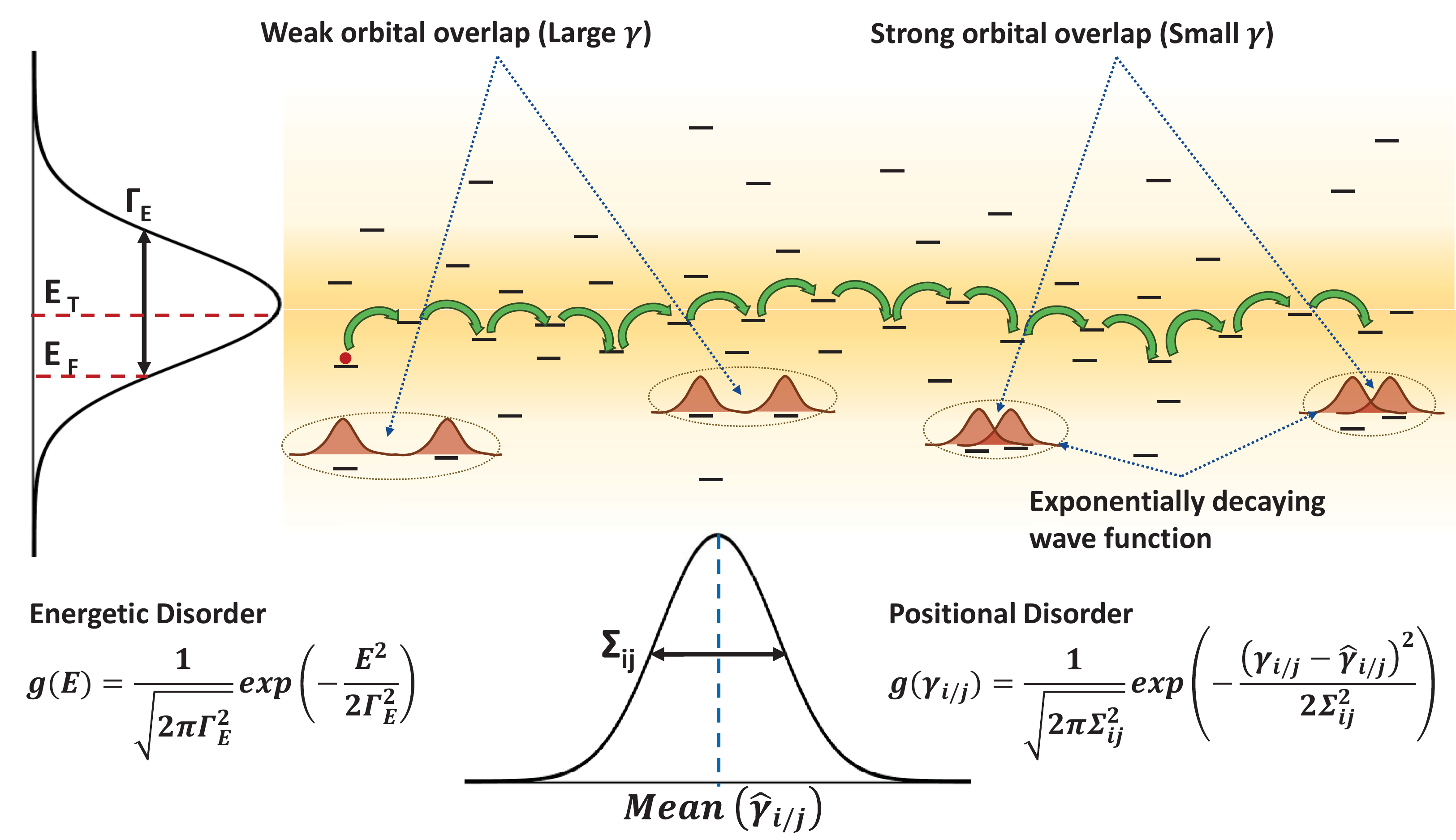}
	\caption{Schematic representation of the carrier hopping process, showing overlap, energetic, and positional disorder.}\label{fig:1}
\end{figure}

Conjugated polymers are positionally disordered systems in which the polymer chains typically interact through weak van der Waals (vdW) interactions. The charge transport within a chain occurs through the covalent framework, and between chains the interactions is through the pi-pi orbitals orthogonal to the chain axis. The displacement of the states about the lattice points causes disruption in the overlap of the pi orbital wave functions termed `positional disorder' \cite{BasslerPSS93}. The interactions between orbitals of adjacent segments are very weak, and the strong electron-phonon coupling in these materials can destroy the coherence between neighboring sites, which causes electrons to become localized to that region and reduces the delocalization range. The decay parameter $\gamma^{-1}$ of the localized electron wave function, or the localization length, is typically 1 $\AA$ $<$ $a\gamma^{-1}$ $<$ 5 $\AA$ \cite{BaranovskiiPRB04}. The vdW and dipole-dipole interactions cause variation in the electrostatic environment \cite{BasslerJCP91}. Furthermore, the dopant molecules Coulombically interact with the carriers localized to a site, thus broadening the density of states (DOS) \cite{ArkhipovPRB05}; this is called `energetic disorder' and shown on the left of Fig.~\ref{fig:1}. The site energies are described by a Gaussian distribution of width $\Gamma_E$ \cite{BasslerJCP91,BasslerPSS93} and the DOS is given as
	\begin{equation}
	g(E) = \frac{1}{\sqrt{2\pi\Gamma_E^2}}exp\left(-\frac{E^2}{2\Gamma_E^2}\right),
	\end{equation}
where $\Gamma_E$ accounts for the degree of energetic disorder in the structure. Positional disorder is modeled as random variations in the wave function overlap parameter $\gamma$ between two sites, as depicted schematically in Fig.~\ref{fig:1}. Hence, $\gamma_{ij} = \gamma_i + \gamma_j$, where  $\gamma_i$ and $\gamma_j$ are the site specific contributions obtained from a Gaussian distribution of width $\Sigma_{ij}$, where the width $\Sigma_{ij}$ accounts for the variation in the electronic wave function coupling due to variation of both the intersite distance and mutual orientation of molecules \cite{BasslerJCP91}.
		
\subsection{Transport Model}

In inorganic semiconductors, ordered crystal lattice with relatively weak electron-phonon coupling leads to a band transport where the interaction between electrons and lattice vibrations (phonons) can be described by perturbation theory. In contrast, polymers do not have long-range periodicity of the atomic structure; electrons are localized and transport is described as a hopping process. The hopping transport process is dependent on temperature \cite{LuPCCP16}. molecular structure, and inter-molecular packing of the material \cite{KiessSpringer92}. Carriers hop from one localized state to another through three possibilities: the electron hops to another state of equal energy by a tunneling process, it hops to a lower energy site while the difference in energy is compensated by the emission of a phonon, or it hops to a state of higher energy and the additional energy required is provided by absorbing a phonon, as illustrated by the green arrows in Fig.~\ref{fig:1}.%, a thermally assisted tunneling process dependent on temperature.

Our model describes the probability that a site is occupied by an electron in terms of Pauli's master equation (PME), which is a differential equation that describes the time rate of change of each site occupation probability due to electrons hopping into and out of it. In the steady-state, the time rate of change of occupation probability will go to zero and PME is given as a sum over all possible transitions into and out of a site
		\begin{equation}\label{eq:pme}
			\frac{dp_i}{dt}= 0 = \sum_{j} [W_{ij} p_i(1 - p_j) - W_{ji} p_j(1 - p_i)],
		\end{equation}
where $p_i$ is the occupation probability of a site $i$ and $W_{ij}$ is the hopping transition rate from site $i$ to $j$, summed over the neighboring sites $j$. The PME is solved for the site occupations using a non-linear iterative solver as described in Methods, after which relevant quantities like mobility and current can be calculated.\cite{VandelholstCh2} The initial site occupation is given by the equilibrium Fermi-Dirac distribution $p_i^0 = \left[\exp\left(\frac{E_i-E_F}{k_BT} \right)+1\right]^{-1}$.

The general hopping rate from site $i$ to site $j$ is given as \cite{FornariJCP15}
		\begin{equation} \label{eq:genrate}
			W_{ij}=\frac{\pi}{\hbar}\sum_{q}^{}\mid M_{ij,q}\mid^2\left[\left(N\left(\omega_q\right)+1\right)\rho_{FCWT}\left(\Delta E_{ij}+\hbar\omega_q\right)+N\left(\omega_q\right)\rho_{FCWT}\left(\Delta E_{ij}-\hbar\omega_q\right)\right] ,
		\end{equation} 
where $\rho_{FCTW}\left(\Delta E\right)$ is a function that depends on the Franck-Condon factors, $\omega_q$is the energy of the phonon mode $q$, $N_q$ is the number of phonons in that mode, given by the Bose-Einstein distribution $N(\omega_q) = \left[\exp\left(\frac{\hbar\omega_q}{k_B T}\right)-1\right]^{-1}$ where $T$ is the absolute temperature. The $M_{ij,q}$ is the phonon-electron coupling constant between sites $i$ and $j$ due to phonon mode $q$, $\Delta E_{ij}= E_j-E_i-eF\Delta R_{ij,x}$ where $E_i$ and $E_j$ are the energies of sites $i$ and $j$ and $F$ is the externally applied electric field. In the limit where there are no phonons with different equilibrium positions in sites $i$ and $j$, such that only transitions $q = q^{\prime}$ can take place, the function $\rho_{FCTW}$ becomes a Dirac delta function and we obtain \cite{WangNL09}
		\begin{equation}
			W_{ij}=\frac{\pi}{\hbar}\sum_{q}^{}\mid M_{ij,q}\mid^2\left[\left(N\left(\omega_q\right)+1\right)\delta\left(\Delta E_{ij}+\hbar\omega_q\right)+N\left(\omega_q\right)\delta\left(\Delta E_{ij}-\hbar\omega_q\right)\right]	.	
		\end{equation}
\noindent Calculation of these rates can be computationally challenging as it requires that we first calculate the electronic wave functions, phonon modes, and the electron-phonon coupling  constants. We can simplify them further by approximating $M_i$ to be proportional to the overlap of the wave functions $\gamma_{ij}=\int d^3r\mid \psi_i(r)\mid\dot \mid \psi_j(r) \mid$, which then yields\cite{MladenovicAFM15}
		\begin{equation}
			W_{ij}=\beta^2\gamma_{ij}^2\left[N\left(\Delta E_{ij}\right)+\frac{1}{2} \pm \frac{1}{2}\right]D_{ph}\left(\Delta E_{ij}\right)/\Delta E_{ij}		
		\end{equation}
where $\beta$ is the coupling constant between electron-phonon coupling constants and wave function overlap, and $D_{ph}(E)$ is the phonon DOS normalized such that $\int_{0}^{\infty}D_{ph}(E)dE=1$. For hops upwards in energy ($E_j>E_i$) by absorption of a phonon, it is $-\frac{1}{2}$ and for downward hops with the emission of a phonon, it is $+\frac{1}{2}$ in the rate equation. Further simplification can be made by assuming that the wave function overlap decays exponentially with distance, and if we ignore the energy dependance we get the Miller-Abrahams rate equation,\cite{MillerPRev60}
		\begin{equation}
			W_{ij} = v_0exp(-2\gamma_{ij}{R_{ij}})\left[N\left(\Delta E_{ij}\right)+\frac{1}{2} \pm \frac{1}{2}\right]
		\end{equation}
where $v_0$ is the attempt-to-escape frequency and $\Delta R_{ij}$ is the distance between the sites. 
	
The Miller-Abrahams rate equation considers only bare charge transport. Since the phonon-electron coupling is strong in organic polymers, it is important to consider the affect of polaron transport, and analyze its effect on TE properties. As a polaron moves through different states, there is deformation of the molecule as the polaron arrives and leaves, and the energy associated with the relaxation of the molecule upon charge transfer is called the binding or reorganization energy. Using $\rho_{FCTW}\left(\Delta E\right)=\sqrt{\frac{1}{4\pi E_0k_BT}}exp\left[-\frac{\left(\Delta E+E_0\right)^2}{4E_0k_BT}\right]$,\cite{FornariJCP15} where $E_0$ is the reorganization energy in Eq. \ref{eq:genrate}, and further simplifying we get
		\begin{equation}
			W_{ij} = v_0\sqrt{\frac{1}{4\pi E_0k_BT}}exp\left(-2\gamma_{ij}R_{ij}\right)exp\left[-\frac{\left(-\Delta E_{ij}+E_0\right)^{2}}{4E_0k_bT}\right],
		\end{equation}
which is the Marcus rate equation.\cite{MarcusRPC64} It is important to note that we can also obtain the Miller-Abrahams rate by taking the limit $E_0\to 0$ in the Marcus rate.
	
The non-linear PME is solved using these rates on a 35$\times$25$\times$25 lattice of sites with an average distance between adjacent lattice points $a$=1 nm. The number of \textquoteleft neighbors\textquoteright depends on the hopping distance and lattice described; we have a cubic lattice and consider up to the third-nearest neighbor, which implies hopping to the nearest 26 sites and maximum hopping distance of $\sqrt{3a}$. The electronic wavefunctions are localized so their overlap has an exponential decay with distance; hence the probability of hopping to neighbors further than the third-nearest neighbor is very small and does not contribute as significantly to transport. The $(1-p_{i/j})$ factor accounts for the exclusion principle requiring that only one carrier is occupying a particular site, due to the high Coulomb penalty for the presence of two charges on one site. All simulations are run under low-field conditions with field $F$=10$^6$ Vm$^{-1}$, attempt to jump frequency $v_0$=10$^{12}$ s$^{-1}$, energy distribution width $\Gamma_E$=3k$_B$T, overlap $\gamma$=3 and $T$=300 K unless stated otherwise. We obtain the current density $J$ by a summation over all the carriers in the direction of the applied field \cite{PasveerPRL05} (here the x direction)
		\begin{equation}
			J = \frac{e}{a^3N}\sum_{i,j}W_{ij}p_i(1-p_j)R_{ij,x}
		\end{equation}
\noindent where $e$ is the electron charge, and then for conductivity we take $\sigma = J/F$.

The Seebeck coefficient (or thermopower) is the voltage built up in response to an applied temperature gradient, given by $S=-\frac{\Delta V}{\Delta T}|_{I=0}$, as carriers that respond to an electric field can also be elicited by a temperature gradient. While each carrier carries a charge $e$, it also carries an `excess' energy $E - E_F$ \cite{ZvyaginPSSB80,KangNMat17}, and the Seebeck coefficient can be calculated as the average carrier energy   
		\begin{equation}\label{eq:Sb}
			S = \frac{(E_F - E_T)}{eT}   
		\end{equation}
where $E_T$ is the average transport energy, calculated from \cite{TesslerPCL14}		
		\begin{equation}\label{eq:Sb2}
		E_T = <E_i> = \frac{\sum_{i,j}E_i W_{ij}p_i(1-p_j)R_{ij,x}}{\sum_{i,j}W_{ij}p_i(1-p_j)R_{ij,x}},
		\end{equation}
\noindent where the brackets $<.>$ denote an average over the sites. The Lorenz number is related to the open-circuit electronic thermal conductivity \cite{WangJAP18}
\begin{equation}
\kappa_o = \frac{\sum_{i,j}(E_i-E_F)^2 W_{ij}p_i(1-p_j)R_{ij,x}}{eT^2}
\end{equation}
\noindent through $L=\kappa_o/\sigma T - S^2$ and thus can be analogously obtained from the variance of the excess energy \cite{WangJAP18} 
\begin{equation}
L=\left\langle \left(\frac{E_i-E_F}{eT}\right)^2\right\rangle - \left\langle \left(\frac{E_i-E_F}{eT}\right)\right\rangle^2.
\end{equation}
			
\section{Results and Discussion}

	\begin{figure}[h]
		\centering
		\includegraphics[width=0.9\textwidth]{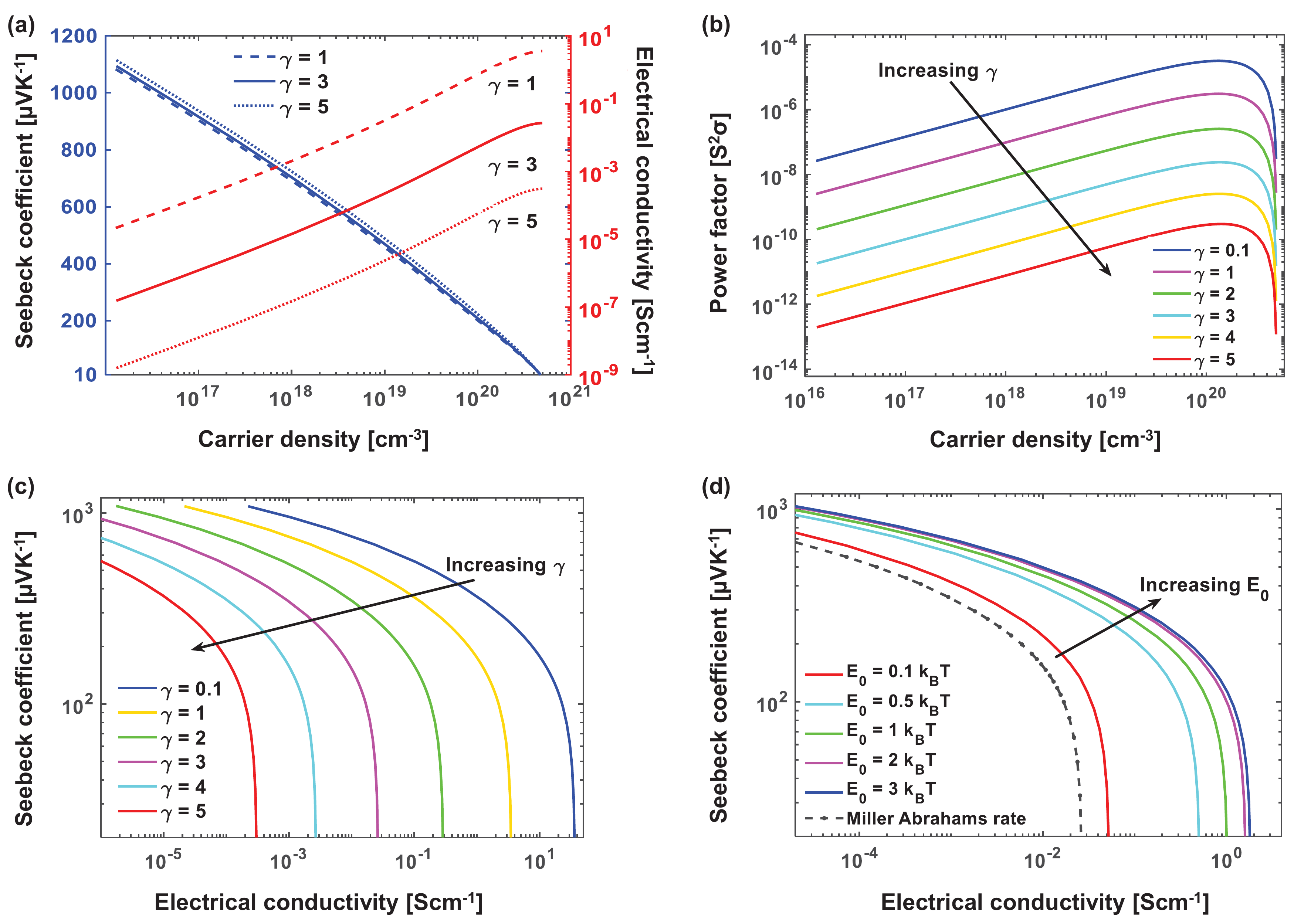}
		\caption{(a) Seebeck coefficient, electrical conductivity and (b) power factor as a function of carrier density at different overlap parameter values. We note than the highest power factor is obtained at a carrier density of $2\times10^{20}$ c$m^{-3}$, corresponding to 20\% of sites being occupied. (c) Seebeck coefficient vs. Electrical conductivity showing the increase in electrical conductivity and Seebeck coefficient with stronger electronic orbital overlap. (d) Comparison of the conductivity vs. Seebeck plot for hopping rates computed using Miller-Abrahams and Marcus rate equations.}\label{fig:2}
	\end{figure}

First, we consider the impact of doping and overlap between neighboring sites on TE properties. Fig. \ref{fig:2}a shows the dependence of electrical conductivity and Seebeck coefficient on the carrier density. We increase the carrier density by moving the Fermi level $E_F$ closer to the center of the Gaussian energy distribution, analogous to electrochemical doping of polymers. We can clearly see the inverse relation that exists between these parameters and the charge density, and the challenge it poses to obtaining high power factor values. At low concentration, the Seebeck is in the range of hundreds of $\mu$VK$^{-1}$, but increasing the carrier concentration causes it to decrease to a few tens of $\mu$VK$^{-1}$. However, the conductivity has an appreciable value only at high concentrations and the highest power factor is achieved at a carrier concentration of $2\times10^{20}$ c$m^{-3}$, shown in Fig. \ref{fig:2}b, which corresponds to 20\% of sites being occupied by carriers. 

We plot Seebeck coefficient vs. conductivity for different values of the overlap parameter in Fig. \ref{fig:2}c, where each point on the curve represents the parameters computed at different carrier densities. The advantage of such a plot is that one can readily see the effect of both carrier density and overlap parameter on Seebeck and conductivity, wherein a curve bulging more towards the top right indicates higher power factor. We find that electrical conductivity is strongly dependent on the overlap parameter whereas it has negligible effect on the Seebeck coefficient, as seen in Fig. \ref{fig:2}a. A smaller value of the overlap parameter, which implies stronger electronic orbital overlap between adjacent sites, is favorable for the hopping process, and thus increases the electrical conductivity. In Fig. \ref{fig:2}d we compare the Seebeck and conductivity computed from Miller-Abrahams and Marcus rate equations. The curve shifts right for higher values of reorganization energy $E_0$, showing improved conductivity with increasing polaronic binding, while as we decrease $E_0$ and approach the limit $E_0\to 0$, it falls back to the curve obtained from Miller-Abrahams rate.  
	
	\begin{figure}[h]
		\centering
		\includegraphics[width=0.9\textwidth]{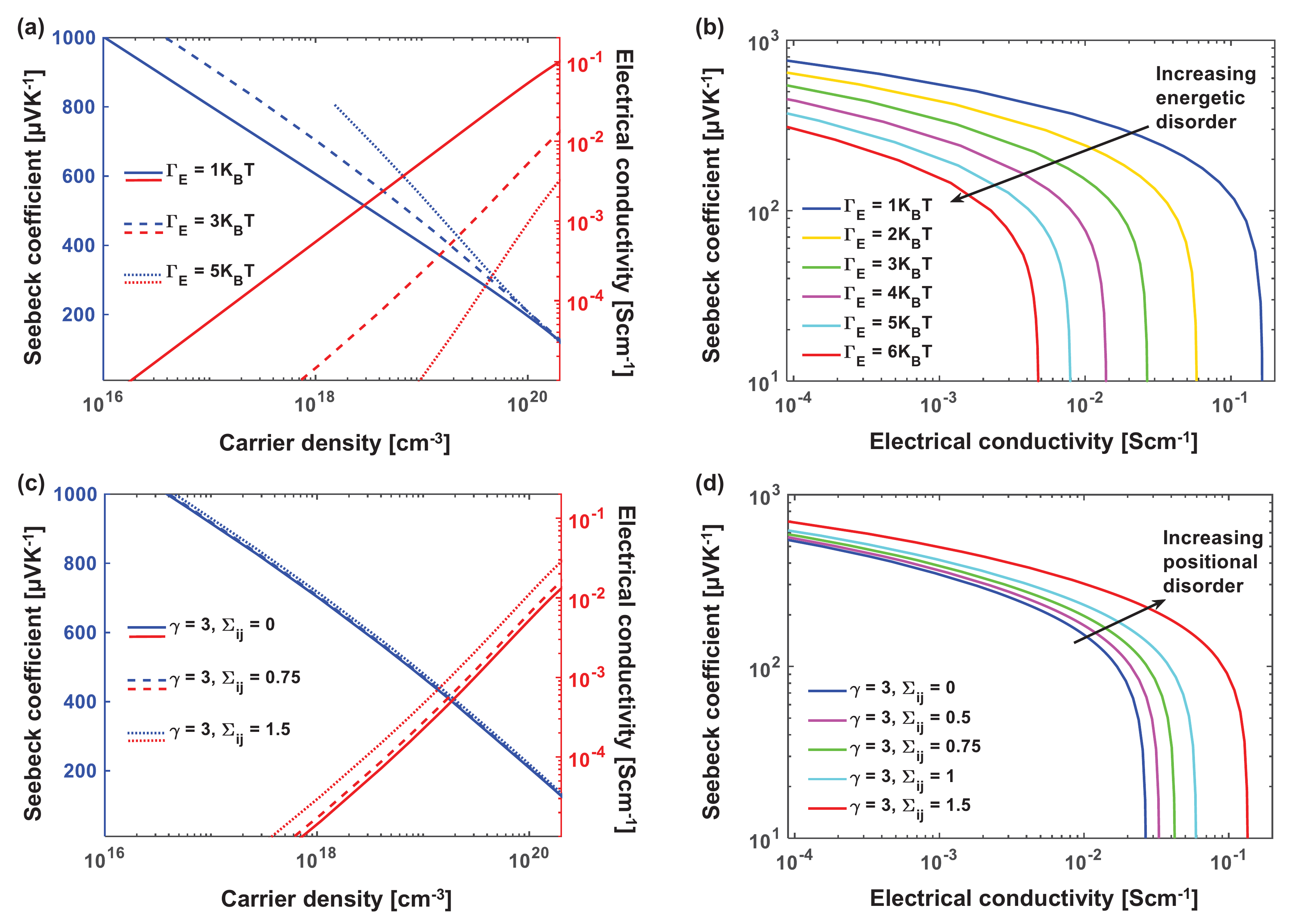}
		\caption{Seebeck coefficient and electrical conductivity vs. carrier density for varying widths of (a) Gaussian energy distribution (energetic disorder), and (c) distribution of the overlap parameter (positional disorder). Seebeck coefficient vs. electrical conductivity with varying (b) energetic and (d) positional disorder. We find that smaller energetic disorder and larger positional disorder lead to better TE performance of a material.}\label{fig:3}
	\end{figure}

Next, we explore the effect of varying degrees of energetic and positional disorder in the system on its TE performance. A larger variation of the Gaussian energy distribution ($\Gamma_E$), leads to a decrease in conductivity and small increase in Seebeck (see Fig. \ref{fig:3}a). Decreasing the width of the energy distribution, meaning a more sharply peaked DOS, lowers the spread in the site energies, leading to a smaller difference $\Delta E_{ij}$ between energies of neighboring sites. A favorable network of nearly equal energy sites thus forms, alleviating the required thermal assistance by absorption of phonons. It is widely thought, based on the work on Mahan and Sofo \cite{MahanPNAS96} that a narrower DOS leads to a higher thermoelectric figure-of-merit ZT, with a delta-function DOS being ideal. 

Combining Eqns. \ref{eq:Sb} and \ref{eq:Sb2} we see that the Seebeck coefficient can be viewed as the average “excess” (away from the Fermi level) entropy per carrier $S=<E_F-E_i>/eT$. Increasing energetic disorder broadens the DOS and makes it flatter but does not affect the shape of the Fermi-Dirac distribution function. Consequently, there are nearly as many states near the DOS peak as there are away from it, which pushes the average $E_i$ away from the center of the DOS. Reducing energetic disorder has the opposite effect: the DOS becomes more sharply peaked with many more states near the peak, resulting in an average $E_i$ closer to the middle of the DOS. If we fix the Fermi level $E_F$ and compare, then larger energetic disorder would imply smaller Seebeck and vice versa. However, if we compare while keeping the carrier concentration $n(E_F,T)=\int{g(E)p_i^0(E,T)dE}$ constant, then the opposite trend emerges: increasing energetic disorder pushes the Fermi level away from the center of the DOS, countering the change in $E_i$ and thus slightly increasing the Seebeck, shown on the left axis in Fig. \ref{fig:3}a. More quantitatively, based on the Mott formula for the Seebeck coefficient,\cite{HeremansSci08} $S=-(\pi^2/3)(k^2T/q)\frac{\partial}{\partial E}ln[\sigma(E)]|_{E=E_F}$ for carriers near the Fermi level, which can be expanded for hopping transport as $S\propto\frac{dln[g(E)]}{dE}+g\frac{d[\mu(E)]}{dn}$.\cite{IhnatsenkaPRB15} In the case of GDM and Gaussian DOS, the first term is dominant and becomes $-E_F/\Gamma_E^2$, while the second term is typically small. Overall, Fig. \ref{fig:3}b shows that the effect of disorder on conductivity is more pronounced and reducing energetic disorder leads to higher TE power factors.

Next, we vary the amount of positional disorder in the system by varying the width of the Gaussian overlap distribution ($\Sigma_{ij}$). Positional disorder primarily affects the conductivity, while the Seebeck is mostly sensitive to the distribution of energies (DOS) and less on their relative positions (Fig. \ref{fig:3}c). The conductivity increases with increasing positional disorder, and shifts the curves right in Fig. \ref{fig:3}d, signifying higher power factor values. This is due to the increase in overlap of approximately half the near-neighboring sites in the system, which aids the formation of conduction paths, consequently increasing the probability of hopping into higher-energy sites. Thus, larger positional disorder but smaller energetic disorder are desirable for polymer TEs.	

	\begin{figure}[h]
		\centering
		\includegraphics[width=0.9\textwidth]{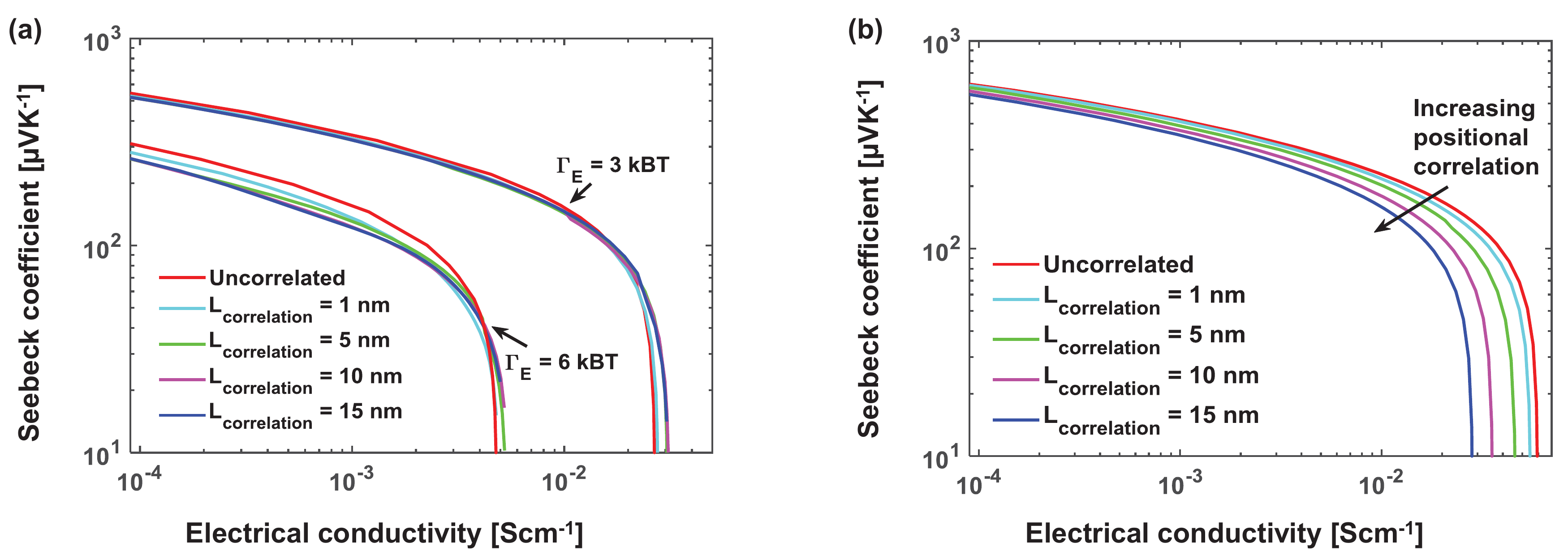}
		\caption{Seebeck coefficient vs. electrical conductivity for different degrees of correlation in the (a) Gaussian energy distribution, (b) distribution of the overlap parameter. We find that correlation in energy distribution has negligible affect, whereas a modest improvement in conductivity is observed with smaller correlation in the orbital overlap distribution.}\label{fig:4}
	\end{figure}
	
In the GDM, site energies are distributed independently with no correlations occurring over any length scale. Nonetheless, spatial fluctuations and corresponding correlation of energy arising from charge dipole interactions and molecular density fluctuations should affect transport \cite{NovikovPRL98}. It has been shown that energy correlation leads to Poole-Frenkel field dependence of mobility over a wide range of fields \cite{GartsteinCPL95,NovikovPRL98}. Spatial correlation in the energetic landscape, modeled by averaging energy over neighboring sites, was also shown to lower the transport energy and decrease the Seebeck coefficient \cite{TesslerPCL14}. However, the impact of correlation length has not been firmly established in the context of TE properties, nor has the effect of correlation on positional disorder through the overlap parameter been studied. 

To further explore the effect of long-range correlation on TE parameters, we use an inverse fast Fourier transform (IFFT) method \cite{WuTB00, BuranIEEE09, RamayyaPRB12} to generate autocorrelated distributions of energy and overlap parameter with a specific correlation length. We start from the standard exponential form of the autocorrelation function of the site energies $\mathcal{C}(R_{ij})=<E(R_i)E(R_j)>=\Gamma_{E}^2 \exp\left(-\sqrt{2}R_{ij}/L_{corr}\right)$ where $R_{i/j}$ are the positions of the $i/j$'th sites, $R_{ij}$ is the distance between the two sites, and $L_{corr}$ is the correlation length. The spectral density is Fourier transform of the autocorrelation function $|\mathcal{S}|^2=\mathcal{F}(\mathcal{C})$. The autocorrelated distribution is obtained by multiplying a random phase ($e^{i\phi}$), having angle $\phi$ uniformly distributed between (0, 2$\pi$), with the square root of spectral density and then taking the inverse Fourier transform \cite{MaurerAPL15}
		\begin{equation}
			\textit{E}=\mathcal{F}^{-1}\left(|\mathcal{S}|e^{i\phi}\right).
		\end{equation}
\noindent The same procedure is applied to obtain a spatially-correlated distribution of overlap parameters $\gamma$. This method allows us to vary the correlation length independently of the variance. Fig. \ref{fig:4}a shows the Seebeck coefficient vs. electrical conductivity for uncorrelated and correlated energy distribution with different correlation lengths and fixed $\Gamma_{E}$=3 and 6 k$_B$T. The same is shown in Fig. \ref{fig:4}b for correlation in overlap distribution. We find that energy correlation has virtually no effect on the TE parameters, while the curve shifts left with increasing correlation length for the overlap distribution. We conclude that smaller correlation between sites leads to better conductivity but the Seebeck remains largely unaffected, thus effectively decoupling these parameters.
	
		\begin{figure}[h]
			\centering
			\includegraphics[width=\textwidth]{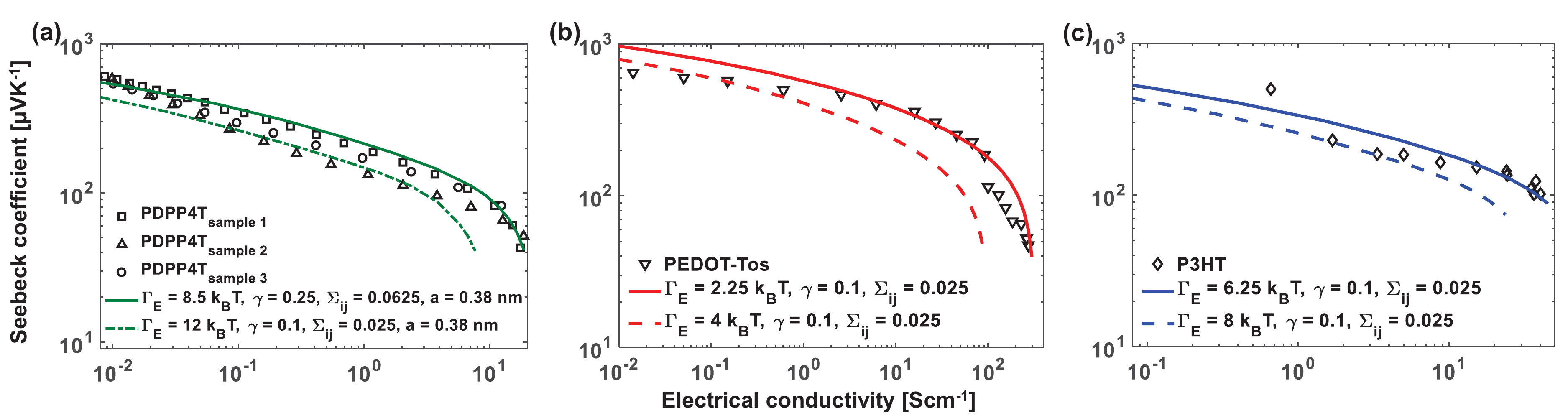}
			\caption{Comparison of our model to experimental data from (a) PDPP4T samples from our measurements, (b) PEDOT:Tos from Ref.~\onlinecite{BubnovaNmat11}, and (c) P3HT sample from Ref.~\onlinecite{JungSciRep17}, showing good agreement across multiple data sets. We plot two curves (solid and dashed lines) on the top and botton of the experimental data to show that the values would fall between these two extremes and account for possible error bars.}\label{fig:5}
		\end{figure}
	
We compare our model to experimental data from several measurements, including three PDPP4T samples (Fig. \ref{fig:5}a) that we chemically doped with iodine and measured as described in Methods, PEDOT:Tos from Ref.~\onlinecite{BubnovaNmat11} (Fig. \ref{fig:5}b), and P3HT sample from Ref.~\onlinecite{JungSciRep17} (Fig. \ref{fig:5}c). Our results are in good agreement with the data and the fit is obtained by varying relevant parameters including overlap, average distance between adjacent lattice points, energetic, and positional disorder. The data for PEDOT:Tos reported to have a ZT$\sim$0.25 obtained from Ref.~\onlinecite{BubnovaNmat11}, is a close fit to the $\gamma$=0.1 and $\Gamma_E$=2.25 $k_BT$ curve, implying stronger electronic orbital overlap between adjacent sites and small energetic disorder, which explains the exceptional conductivity observed in these samples beyond what is obtained from $\gamma$=3, a value commonly used in calculations.
	
\begin{figure}[h]
			\centering
			\includegraphics[width=0.9\textwidth]{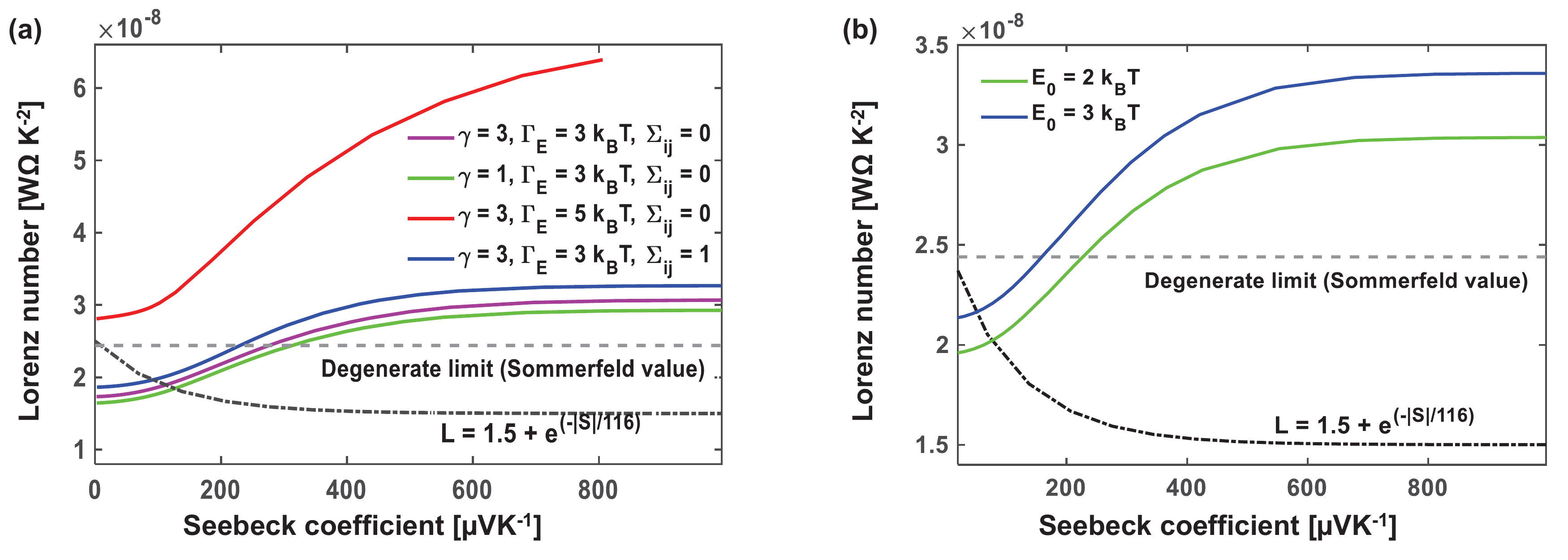}
			\caption{Lorentz number vs. Seebeck coefficient showing the increasing deviation from the Sommerfeld value with increasing (a) overlap parameter, energetic and positional disorder, and (b) polaronic binding energy ($E_0$).}\label{fig:6}
		\end{figure}

Lastly, we turn our attention to the Lorenz number. In most materials, the Lorenz number ranges from a value close to the Sommerfeld value found in metals and degenerate semiconductors $L_0=\pi^2/3 (k_B/e)^2 = 2.44\times 10^{-8}$ W$\Omega$ K$^{-2}$ \cite{Sommerfeld27} decreasing to the value $L_0=2 (k_B/e)^2 = 1.49\times10^{-8}$ W$\Omega$ K$^{-2}$ found in single-parabolic-band materials when acoustic phonon scattering is dominant \cite{ThesbergPRB17}. It has been shown that a first-order correction to the degenerate limit $L=1.45+exp(-|S|/116)$ (where L is in 10$^{-8}$ W$\Omega$K$^{-2}$ and S in $\mu$VK$^{-1}$) is a good approximation and holds across all known band semiconductors \cite{KimAPL15}. In contrast, we see an opposite trend in organic semiconductors, where $L$ increases with Seebeck coefficient as shown in Fig. \ref{fig:6}. Increasing the overlap, positional disorder and polaronic binding energy increases the value of Lorenz number further, but the largest impact is seen with energetic disorder when $\Gamma_{E}$ is increased from 3 to 5 k$_B$T. Experimental data also confirm L is significantly larger than $L_0$ in PEDOT:Tos ~\cite{BubnovaNmat11}.% however Lu et al.\cite{NianduanJAP16}, show a decrease in Lorenz number with increasing energetic disorder.
In the limit when electronic thermal conductivity dominates ($\kappa_e>\kappa_{l}$), the ZT goes to $ZT=S^2/L$; therefore, a simultaneous increase in S and L could have a negative impact on ZT and hence design of effective TE materials with polymers requires consideration of the Lorentz number as well, carefully balancing the roles of disorder and correlation.
	
\section{Conclusion}
	
Polymers, with their inherently low thermal conductivity and low cost of manufacturing, are a promising choice for TE applications. However, for commercial success the figure of merit of these materials needs to be improved further. We study the effects of disorder and correlation on their TE properties using a hopping transport model. We find that positional disorder leads to a moderate increase in the electrical conductivity whereas the Seebeck remains unaffected. Energetic disorder has an adverse affect on conductivity but leads to a moderate increase in Seebeck coefficient at lower doping concentrations. Consequently, positional correlation primarily affects conductivity, while correlating the nearby site energies has no effect on the TE properties. We conclude that controlling energetic and positional disorder would allow us to decouple conductivity and the Seebeck coefficient. Minimizing energetic disorder and correlation while increasing positional disorder results in a higher TE power factor. Our results also show that the Lorenz number increases with the Seebeck coefficient, more so with increasing disorder, in contrast to the universal trend observed across all materials exhibiting band transport. %Disorder provides additional avenues to tune the TE properties in polymers, however it is important to find the right combination of doping and disorder, and their effect of the conductivity, Seebeck and Lorenz number for effective design of TE materials. 
We find that numerical transport models can play a key role in predicting the optimum structural characteristics and aid the design and development of novel materials for TE applications. 
	
\section{Methods}

\subsection{Solving the non-linear PME}

We solve the non-linear PME using a standard iterative non-linear solver. First, we cast the PME as zero-finding for a system of equations $F_i(\mathbf{p}) = \sum_{j} [W_{ij} p_i(1 - p_j) - W_{ji} p_j(1 - p_i)] = 0$, which can be written in terms of the in- and out-scattering as $F_i(\mathbf{p}) = p_i S_{out}(\mathbf{p}) - (1-p_i)S_{in}(\mathbf{p})$ with $S_{out}(\mathbf{p})=\sum_{j} W_{ij} (1 - p_j)$ being the out-scattering term and $S_{in}(\mathbf{p})=\sum_{j} W_{ji} p_j$ being the in-scattering. The expression for $F_i(\mathbf{p})$ is nonlinear because both in- and out-scattering terms depend on the unknown $\mathbf{p}$. Previous studies typically rearranged this equation to obtain a fixed point iteration for the $p_i$ such as $p_i^{n+1} = S_{in}(\mathbf{p}^n)\left/ [S_{in}(\mathbf{p}^n)+S_{out}(\mathbf{p}^n)] \right.$ with the initial condition $p_i^0$ being the equilibrium Fermi-Dirac distribution given earlier. However, a fixed point iteration can stall, resulting in poor convergence for some cases. For this reason, we follow the same fixed-point iteration procedure here for the first few iterations and then use the resulting estimate of $p_i$ as an initial guess in the next step where we numerically solve for the zero of $F_i(\mathbf{p})$. Rather than solving for the site occupancies $p_i$, we solve for their deviation away from equilibrium $\Delta p_i=p_i-p_i^0$. %In equilibrium, the principle of detailed balance requires that $W_{ij} p_i^0(1 - p_j^0) - W_{ji} p_j^0(1 - p_i^0) = 0$. 
Combining this with Eq. \ref{eq:pme} and simplifying for $\Delta p_i$, we get $F_i(\mathbf{p})= \Delta p_i S_{out}(\mathbf{p}) - (1 - p_i^0) S_{in}(\mathbf{p}) = 0$. We arrange the 35$\times$25$\times$25 array of $\Delta p_i$'s into a column vector $\mathbf{p}$ and compute the Jacobian matrix of derivatives of $F_i$ with respect to $p_j$ as $J_{ij}=dF_i/dp_j = -W_{ij} p_i - W_{ji} (1- p_i)$. Then we apply the Levenberg-Marquardt algorithm,\cite{NumRecBook} as implemented in MATLAB's fsolve function, with the known Jacobian matrix, which requires a linear solve at each iteration but typically converges in a few iterations due to its high rate of convergence. The linear solver is a preconditioned Conjugate Gradients algorithm with a banded preconditioner based on an incomplete Cholesky factorization.

\subsection{PDPP4T sample preparation and characterization}

PDPP4T (molecular weight M$_n$=72.8 kDa and dispersity \DJ=4.4)  was synthesized according to previously reported procedures \cite{LiACS11}. \textbf{Film preparation.} A solution 8 mg/mL PDPP4T in chloroform was stirred for no less than 4 h prior to dropcasting 0.23 mL of the polymer solution onto a handcut, 1.1 cm $\times$ 2.2 cm glass coverslip that was preheated to 45 $^{\circ}$C on a hot plate. This was immediately covered with a watch glass to impede the escape of chloroform vapors and slow the rate of evaporation during dropcasting. After 10 min, the sample heating element was turned off and the sample was let stand under ambient conditions for no less than 24 h to further evaporate the chloroform. \textbf{Doping with iodine vapor.} 50$\pm$5 mg iodine crystals were loaded into a 1 mL glass vial, and this vial was loaded into a 20 glass mL vial and sealed with a cap closed for no less than 12 h to allow iodine vapor to diffuse within the vial. Iodine doping was carried out by placing the polymer film into this 20 mL vial, immediately resealing by capping the vial, and letting stand for 24 h.

\subsection{Characterization of thermoelectric properties}

Samples were transferred from their iodine doping chamber to a custom-built apparatus for thermoelectric characterization in a timely fashion, since they began dedoping immediately and rapidly in the absence of iodine vapor. The details of this apparatus are reported elsewhere \cite{TongaAMI17}. In brief, a thermal gradient is established between two copper blocks by heating one of them and leaving the other at ambient temperature. The polymer film was placed onto a glass slide located on these copper blocks, and Pt wire electrodes embedded in a PTFE block were firmly secured onto the sample. A LabView program was used to interface with a digital dual input thermometer with k-type thermocouples, a Keithley 6182 nanovoltmeter, and a Keithley 2440 source meter to repeat measurements of the temperature gradient $\Delta T$, voltage gradient $\Delta V$, and I-V characteristics across the sample sequentially and every 2 min. $\Delta V$ was taken to be the average of 1000 voltage measurements from the Keithley 6182 nanovoltmeter and the Seebeck coefficient was taken to be $\Delta V/\Delta T$. Only measurements for which the standard deviation of the 1000 voltage measurements is less than 1\% of their mean are considered to calculate the Seebeck coefficient and the rest discarded to ensure only reliable estimates are used to compare to our model. The conductance was taken to be the slope of the $I-V$ curve and electrical conductivity was taken to be $G\times l/A$. $l$ is the length between two platinum electrodes and $A$ is the area of the device and is taken to be $w \times t$ where $w$ is the width of the film and $t$ is the thickness obtained from profilometry after thermoelectric measurements were finished. Since the iodine doped polymer film is initially heavily doped, and spontaneously dedopes over time, these repeated measurements collect Seebeck coefficients and conductivities across a broad range of doping. We will describe this methodology in more detail in an upcoming report.
	
\section{Data Availability.} The datasets and codes generated and/or analysed during the current study are available from the corresponding author on reasonable request.

\newpage
\makeatletter
\renewcommand\@biblabel[1]{#1.}
\makeatother
\bibliographystyle{naturemag}
\bibliography{thermoelectric}	

\section{Acknowledgements}
The authors thank Dr. Feng Liu and Prof. Thomas P. Russel of the University of Massachusetts Amherst for providing PDPP4T for these studies. The authors also thank Profs. Frank E. Karasz and Paul M. Lahti for the thermoelectric measurement equipment and Dr. Ljiljana Korugic for useful discussions.

\section{Author Contributions}

Z.A. and D. V. conceived the idea. Z. A. implemented the simulation code. M. U. performed the simulations and created the graphs. D. V. and C. J. B. designed and implemented the experimental setup. C. J. B. performed the measurements. All authors co-wrote the manuscript. 

\section{Competing interests}
The authors declare no competing interests.
	
\end{document}